\documentclass[aps,prb,preprint,amsmath,amssymb,superscriptaddress,floatfix,scrartcl]{revtex4-1}

\usepackage{amssymb}
\usepackage{color}
\usepackage{graphicx}
\usepackage{srcltx}
\usepackage{mathrsfs}
\usepackage[pdftex,pdfusetitle,
 bookmarks=true,colorlinks=true,citecolor=blue,urlcolor=blue,linkcolor=red]{hyperref}
\usepackage{hypcap}

\begin{document}

\title{
Out-of-time-order correlations in  many-body localized  and  thermal phases
   }

\author{Xiao Chen}
\email{xchen@kitp.ucsb.edu}
\affiliation{Department of Physics and Institute for Condensed Matter Theory, University of Illinois at Urbana-Champaign, 1110 West Green Street, Urbana, IL 61801-3080, USA}
\affiliation{Kavli Institute for Theoretical Physics, University of California at Santa Barbara, CA 93106, USA}

\author{Tianci Zhou}
\email{tzhou13@illinois.edu}
\affiliation{Department of Physics and Institute for Condensed Matter Theory, University of Illinois at Urbana-Champaign, 1110 West Green Street, Urbana, IL 61801-3080, USA}

\author{David A. Huse}
\email{huse@princeton.edu}
\affiliation{Department of Physics, Princeton University, Princeton, NJ 08544, USA}
 
\author{Eduardo Fradkin}
\email{efradkin@illinois.edu}
\affiliation{Department of Physics and Institute for Condensed Matter Theory, University of Illinois at Urbana-Champaign, 1110 West Green Street, Urbana, IL 61801-3080, USA}

\date{\today}

\begin{abstract}
We use the out-of-time-order (OTO) correlators to study the slow dynamics in the many-body localized (MBL) phase.  We investigate OTO correlators in the effective (``l-bit'') model of the MBL phase, and show that their amplitudes after disorder averaging approach their long-time limits as power-laws of time.
This power-law dynamics is due to dephasing caused by interactions between the localized operators that fall off exponentially with distance.
The long-time limits of the OTO correlators are determined by the overlaps of the local operators with the conserved l-bits.  We  demonstrate numerically our results in the effective model and three other more ``realistic'' spin chain models. Furthermore, we extend our calculations to the thermal phase and find that for a time-independent Hamiltonian, the OTO correlators also appear to vanish as a power law at long time, perhaps due to coupling to conserved densities.  In contrast, we find that in the thermal phase of a Floquet spin model with no conserved densities the OTO correlator decays exponentially at long times.
\end{abstract}


\maketitle

\section{Introduction}
Quantum thermalization is a topic that has received renewed attention in different areas of physics. For a realistic isolated quantum many-body system totally decoupled from the environment, the interactions can produce chaos and at long times thermalize all small subsystems under the dynamics of the full isolated system.  Such systems then act as their own bath and are in a thermal phase.\cite{Deutsch1991,Srednicki1994}

On the other hand, not all interacting systems succeed in thermalizing themselves. One fascinating example is a many-body localized (MBL) phase where quenched disorder prevents thermalization.\cite{Basko06,Oganesyan07,Pal2010,Serbyn2013b, Huse2014} The MBL phase is an interacting many-body generalization of Anderson localization.\cite{Anderson-1958} Deep inside the MBL phase with strong disorder, the Hamiltonian can be effectively described in terms of a set of conserved and localized integrals of motion (sometimes called ``l-bits'') with exponentially decaying interactions between them.\cite{Serbyn2013b,Huse2014} These weak long-range interactions are responsible for slow long-time dynamics observed in the MBL phase which are absent in noninteracting Anderson localization.\cite{Znidaric2008, Bardarson2012, Vosk2013, Serbyn2013a, Serbyn2014a, Serbyn2014b, Huse2014} For a review of MBL phases and their comparision with thermal phases and noninteracting localized phases, see Ref. \onlinecite{Nandkishore2015} and Table 1 therein.

Recently, the out-of-time-order (OTO) correlator has been proposed to measure the dynamics of quantum chaos and entanglement in many-body quantum systems.\cite{Larkin1969, Shenker2013a, Shenker2013b, Shenker2014, Maldacena2015, Kitaev2015, Hosur2015,Roberts2016} The OTO correlator of two Hermitian operators
for a given pure state is defined as
\begin{align}
O_s(r_1,r_2;t)=\langle \psi|\hat{W}(r_1,t)\hat{V}(r_2,0)\hat{W}(r_1,t)\hat{V}(r_2,0)|\psi\rangle ~,
\end{align}
where $|\psi\rangle$ is some initial state.  The behavior is generally qualitatively the same for $\pm t$ due to the reversibility of the dynamics, so we will for specificity and familiarity choose to look at positive time $t$.  The same quantity can also be defined for a thermal average
\begin{align}
O_{\beta}(r_1,r_2;t)=\langle \hat{W}(r_1,t)\hat{V}(r_2,0)\hat{W}(r_1,t)\hat{V}(r_2,0)\rangle_{\beta} ~,
\end{align}
where $\langle\cdot\rangle_{\beta}=\mbox{Tr}[e^{-\beta H}\cdot]/Z$ represents the usual thermal expectation value and $\hat{W}(r_1,t)=e^{i\hat{H}t}\hat{W}(r_1,0)e^{-i\hat{H}t}$ is the time evolution of a local operator $\hat{W}(r_1,0)$. At $t=0$, $\hat W(r_1,0)$ and $\hat V(r_2,0)$ are two local operators inserted at locations $r_1$ and $r_2$ separated by a distance $r=|r_1-r_2|$.  For $r\neq 0$ and $t=0$ they commute.  Therefore, $O_{\beta}(r_1\neq r_2; t=0)=\langle\hat W^2\hat V^2\rangle_{\beta}$.

As time goes on, the evolved operator $\hat W(r_1,t)$ spreads in space and its commutator with $\hat V(r_2,0)$ becomes nonzero and grows.  For simplicity, we can consider the single state OTO correlator $O_s(t,r)$, which can be regarded as the overlap of the two states
\begin{equation}
|1 \rangle = \hat{W}(r_1,t)\hat{V}(r_2,0)|\psi \rangle \quad |2 \rangle =  \hat{V}(r_2,0)\hat{W}(r_1,t)|\psi \rangle ~.
\end{equation}
In a spacetime diagram, the state $| 1\rangle$ corresponds to the following four operations applied on state $|\psi\rangle$: (1) acting with the $V$ operator at $r_2$, (2) evolving to time $t$, (3) acting the $W$ operator at $t$ and $r_1$, (4) finally doing backward time evolution to the initial time. On the other hand, the state $|2 \rangle$ is produced by having the $V$ operator act last.  
In a chaotic system, this difference will change the state significantly and therefore indicates the strength of the ``butterfly effect''. Since we expect the overlap to be smaller as time goes on, the OTO correlator for a single state in general will decay. The thermally averaged version is the Boltzmann average over all the energy eigenstates.

In a thermal phase in which chaos is present, the OTO correlator will eventually decay to zero. The precise form of this asymptotic behavior is model-dependent as is discussed in this paper. For the strongly chaotic system, at intermediate times, the OTO correlator is expected to take the universal form \cite{Kitaev2015, Maldacena2015,Roberts2015}
\begin{equation}
O(t,r)\sim c_0-c_1e^{\lambda_L(t-r/v_B)}
\end{equation}
where $v_B$ is some characteristic butterfly velocity that defines the effective ``light cone'' for the spreading of the operators.\cite{Roberts2014,Shenker2013a} This form has been derived using large-$N$ or semiclassical \cite{aleiner} limits and it is not to our knowledge known to what extent it applies beyond these limits.  
In this intermediate time regime, the OTO correlator 
grows exponentially with rate $\lambda_L$, which is the quantum analog of a Lyapunov exponent. 
It is argued to obey an upper bound \cite{Maldacena2015}
\begin{equation}
\lambda_L\leq \frac{2\pi}{\beta}
\end{equation}
Among models with such a thermal phase, the Sachdev-Ye-Kitaev model is one of the fastest scramblers and saturates this upper bound.\cite{Sachdev1993,Kitaev2015,Polchinski2016,Maldacena2016}

In this paper, we will use OTO correlator to study the spread of operators in a MBL phase in one-dimensional systems.  Since the MBL phase has no chaos in it, the OTO correlator need not decay to zero at long times, and we show that the thermally averaged OTO correlator approaches its long time value
as a power-law in time, due to the dephasing caused by the exponentially small effective interactions between two remote regions.  We show how the exponent of this  power-law-in-time decay is related to the entropy density of the dephased state and the decay length of the effective interactions.  Actually, these small interactions are  essential for the slow dynamics observed in the MBL phase, such as the logarithmic growth of entanglement from an initial non-entangled state.\cite{Znidaric2008, Bardarson2012, Vosk2013, Serbyn2013a, Serbyn2014a, Huse2014}  Notice that the OTO correlator can measure the non-commutativity of two local perturbations at different times, therefore it is very similar to the non-local spin-echo protocol (introduced in Ref. \onlinecite{Serbyn2014b}), which detects the influence of local perturbations in one region on another region and probes the long-range interaction effects in a MBL phase.  For a single energy eigenstate, the OTO correlator does not converge at long times, instead it oscillates quasiperiodically forever.  The dephasing only comes when the state contains many different eigenstates that each oscillate differently.

We verify this power law decay numerically in the effective (``l-bit'') model of the MBL phase and three other one-dimensional and more ``realistic'' spin-chain models. These models include an Ising model in a random transverse field and uniform longitudinal field, the Heisenberg model in a random field, and a Floquet (periodically driven) spin model with disorder. In all of these models, we observe the power-law decay of the OTO correlator in their MBL phases. This is different from the thermal phase, where the late time behavior appears to be more model dependent.  In the thermal phase of the Heisenberg model in a random field we find that the OTO correlator decays faster, but still as a power law at late times. We attribute this behavior to coupling to the conserved energy and spin densities. In contrast, in the thermal phase of the Floquet system without any conserved densities, we find that the OTO correlator decays exponentially to zero at long times.

\textit{Note}: Recently, several papers studying OTO correlators in MBL phases were posted on arXiv.\cite{Huang2016, Fan2016, Chen_Y2016, He2016, Swingle2016} 
Here we briefly summarize their results and note where our work differs.  Most of the authors studied the early time behavior when the two operators do not yet strongly overlap in space. For example, in Ref.\ \onlinecite{Huang2016}, the authors discussed the early time behavior of OTO correlators to detect the logarithmic light cone effect.  Ref.\ \onlinecite{Fan2016} showed that OTO correlator deviates from unity as a power law rather than exponentially at early time. They also {\textcolor{blue}derived an interesting theorem} relating the second R\'enyi entanglement entropy to OTO correlators and confirmed with some numerical studies of spin chain models.  Ref.\ \onlinecite{Chen_Y2016} is a generalization of Ref.\ \onlinecite{Hosur2015}. The authors studied the OTO correlator in a doubled system with two operators acting on the input and output channel respectively.  They also focused on the early time behavior and studied the logarithmic light cone effect. Ref.\ \onlinecite{He2016} used the OTO correlator and the related squared commutator to study both the thermal phase and the MBL phase. They observed the large fluctuations of OTO correlators in the MBL phase.  Of these papers, only Ref.\ \onlinecite{Swingle2016} studied the late time behavior when the static operator is well within the light cone of the other. The authors studied OTO correlators and squared commutators in both MBL phases and interacting diffusive metals.  In particular, they investigated the OTO correlator in the effective model of MBL and  showed that OTO correlators decay as power laws, as in our results. The focus of our paper is \emph{the late time behavior} of OTO correlator which corresponds to the regime inside the logarithmic light cone.  We support the power-law decay behavior with extensive numerical calculations on various MBL systems and different decaying behaviors in prototypical quantum chaotic systems.

The rest of paper is organized as follows. In Sec. \ref{sec:eff_model}, we briefly review the effective model of the MBL phase and  investigate analytically the OTO correlators for different local operators in this model. In Sec. \ref{sec:numeric}, we  calculate numerically OTO correlators for several models. In Sec. \ref{sec:num_eff}, we  explore the OTO correlators numerically in the effective model of the MBL phase and verify the analytical predictions of Sec. \ref{sec:eff_model}. In Sec. \ref{sec:trans}, \ref{sec:hei} and \ref{sec:flo}, we further study three one-dimensional spin-$1/2$ models: (B) an Ising model in a random transverse field and uniform longitudinal field, (C) the Heisenberg model with a random field and (D) a Floquet disordered spin chain. In all of these models, we find the power-law behavior of OTO correlators in their MBL phases. In cases (C) and (D), we also extend our calculations to the thermal phase and study how the OTO correlators approach zero at late times. Finally, we summarize in Sec. \ref{sec:con}.

\section{OTO correlator for the effective model of the MBL phase}
\label{sec:eff_model}
\subsection{The effective model of MBL phase}

In this section, we briefly review the effective (``l-bit'') model of the MBL phase.\cite{Serbyn2013b, Huse2014, Imbrie2016} The system we study is a one-dimensional spin-$1/2$ chain. With only short-range interaction and strong disorder, the system can enter a fully many-body localized phase, wherein all the eigenstates of the Hamiltonian are localized. In this phase, the system can be characterized by a complete set of localized pseudospins $\{\tau_i\}$ whose $z$ components are all conserved.  These $\{\tau_i\}$ are sometimes called ``l-bits", and are ``dressed'' spins-$1/2$ constructed from the bare spins $\{\sigma_i\}$.
The effective Hamiltonian of the MBL phase in terms of the $\tau$-operators is
\begin{equation}
\hat{\mathcal{H}}=\sum_i h^0_i\hat{\tau}_i^z+\sum_{i,j}J^0_{ij}\hat{\tau}_i^z\hat{\tau}_j^z+\sum_{ijk}J^0_{ijk}\hat{\tau}_i^z\hat{\tau}_j^z\hat{\tau}_k^z+\ldots~.
\label{eff_H}
\end{equation}
It can be exactly solved since all terms commute, i.e., $[\hat{\tau}_i^z,\hat{\tau}_j^z]=[\tau_i^z,H]=0$.

In Eq. \eqref{eff_H}, the $h^0_i$ are random fields acting on each of the dressed $\tau$-spins. The coefficients $J^0$ in the pseudorandom multi-spin interactions have some characteristic strength which falls off exponentially with increasing distance. In the Anderson localized phase with no interactions, all the $J^0$'s vanish and only the fields $h^0_i$ are nonzero.
In the MBL phase, the interactions $J^0$ are essential for the logarithmic-in-time growth of entanglement entropy and other related effects, which are absent in noninteracting Anderson localization.
\cite{Znidaric2008, Bardarson2012, Vosk2013, Serbyn2013a, Serbyn2014a, Serbyn2014b, Huse2014}

If we focus on two l-bits $i$ and $j$, we can write the Hamiltonian \eqref{eff_H} as
\begin{equation}
\hat{\mathcal{H}}=\hat{\mathcal{H}}^{\bar{i}\bar{j}}+\hat{h}_i^{\bar{j}}\hat\tau_i^z+\hat{h}_j^{\bar{i}}\hat\tau_j^z+\hat{J}^{eff}_{ij}\hat\tau_i^z\hat\tau_j^z~,
\label{eff_h}
\end{equation}
where the first term is the sum of all terms in \eqref{eff_H} that do not involve l-bits $i$ and $j$, the second term is the sum of all terms that do involve l-bit $i$ but do not involve l-bit $j$, the third term is {\it vice versa}, and the last term is the sum of all terms that involve both l-bits so produce interactions between these two l-bits.  It is the effective interaction $\hat{J}^{eff}_{ij}$ that we will be most interested in.  In the MBL phase this interaction typically falls off exponentially with the distance $r$ between the two l-bits,
$\hat{J}^{eff}_{ij}\sim J_0\exp{(-r/\zeta)}$, with a decay length $\zeta$ that in general depends on the state of the system \cite{Huse2014}.  Since this effective interaction depends on the values of $\hat\tau_k^z$ for all the other l-bits $k$, it is an operator.  The contribution of l-bit $k$ to the effective interaction typically falls off with distance as $\sim\exp{(-l/\zeta)}$, where $l$ is the distance between the farthest-apart pair of l-bits among $i$, $j$, and $k$.


\subsection{OTO correlator in the MBL phase}

\subsubsection{Either $\hat{W}$ or $\hat{V}$ is the local $\hat{\tau}^z$}

Suppose we take $\hat{W}(r_1,0)$ to be the local integral of motion $\hat{\tau}^z_{r_1}$. This operator will remain invariant under the time evolution,
\begin{equation}
\hat{W}(r_1,t)=\hat{W}(r_1,0)=\hat{\tau}^z_{r_1}
\end{equation}
Thus this time-evolved operator commutes with any arbitrary local operator $\hat{V}(r_2,0)$ at a different location $r_2$.  And since the commutator is zero, it remains zero for all times, so these two operators commute at all pairs of times:
$[\hat{W}(r_1,t_1),\hat{V}(r_2,t_2)]=0$.  Therefore in this case the OTO correlator is a time-independent nonzero constant, so does not decay with time.  The same argument applies if we instead set $\hat{V}=\hat{\tau}^z_{r_2}$.


\subsubsection{$\hat{W}(r_1,0)=\hat{\tau}^x_{r_1}$ and $\hat{V}(r_2,0)=\hat{\tau}^x_{r_2}$}
In this case, the $\hat\tau^x_{r_1}$ operator of the initial $\hat W(r_1, 0)$ will precess around its $z$-axis at a rate that is affected by the value of $\hat\tau^z_{r_2}$ via the effective interaction $\hat{J}^{eff}_{r_1r_2}$.  This gives the OTO correlator a time-dependence.  This is also true if either or both of the operators are $\hat\tau^y$'s.  Putting in the dynamics explicitly using \eqref{eff_h}, it is only the effective interaction term that gives a time dependence, resulting in
\begin{equation}
O(r_1,r_2;t)=\langle\exp{(it4\hat{J}^{eff}_{r_1r_2}\hat\tau^z_{r_1}\hat\tau^z_{r_2})}\rangle ~.
\label{oto_deriv}
\end{equation}
Since the effective interaction falls off exponentially with the distance $|r_1-r_2|$, the OTO crosses over from time independent at large distances to significantly time dependent at shorter distances, with this crossover happening at a distance that grows logarithmically with time: $\sim\zeta\log{(J_0t)}$.  The eigenstates of $\hat{\mathcal{H}}$ are also eigenstates of all the operators in this expression, so for them the OTO correlator $O_s(r_1,r_2;t)$ simply oscillates in time at a frequency set by the effective interaction between those two l-bits.  However, this frequency differs between eigenstates, which results in dephasing of these oscillations for states that are composed of many different eigenstates, as we discuss in more detail below.

\subsubsection{General choice of local operators}
Following the discussion in the previous subsections, we now consider the OTO correlator for general choices of local operators.  We are particularly interested in the
OTO correlator when $\hat W$ and $\hat V$ are bare local spin operators in a realistic spin model.  In the MBL phase we can expand the bare
spin operators $\hat{W}(r_1,0)$ and $\hat{V}(r_2,0)$ in terms of the $\hat{\tau}^{x,y,z}$ operators as
\begin{equation}
\begin{aligned}
&\hat{W}(r_1,0)=a_1\hat{\tau}^z_{r_1}+b_1\hat{\tau}^{x,y}_{r_1}+\ldots,\\
&\hat{V}(r_2,0)=a_2\hat{\tau}^z_{r_2}+b_2\hat{\tau}^{x,y}_{r_2}+\ldots,
\end{aligned}
\label{expa_W_V}
\end{equation}
where $\ldots$ is comprised of sums of tensor products of $\hat\tau^{x,y,z}$'s at nearby sites.  Thus each operator $\hat{W}$, $\hat{V}$ is generically a sum of a part that consists only of products of $\hat\tau^z$'s so commutes with $\hat{\mathcal{H}}$ and gives only time-independent nonzero contributions to the OTO correlator, and a part that contains $\hat\tau^{x,y}$'s so does not commute.  The contributions to the OTO correlator from the nonconserved terms in both operators oscillate in time and have a zero time average, as in the example of the previous subsection.  In general $O(r_1,r_2;t)$ is essentially time independent at small $|t|$ before the operators spread enough to overlap and interact significantly.  The time dependence appears near and beyond $t\sim(1/J_0)\exp{(r/\zeta)}$, where $r=|r_1-r_2|$.  At long times $O(t)$ generically either converges to or oscillates around some nonzero real constant.

For an eigenstate of $\hat{\mathcal{H}}$ the set of oscillating contributions to $O_s(t)$ is discrete and gives a convergent sum.  As a result, for eigenstates the generic long-time behavior of $O_s(t)$ is a quasiperiodic function of time that is different in detail between eigenstates, since the effective interactions are eigenstate- and sample-dependent.  We are also interested in the thermally averaged behavior where these oscillations instead dephase at long times in $O_{\beta}(t)$.  The dephasing comes because different eigenstates give contributions that oscillate in time at different frequencies.  The change in the frequency from flipping a l-bit at distance $l$ from the farthest of the two sites is $\sim J_0\exp{(-l/\zeta)}$, so this will result in dephasing between these two eigenstates after time $t\sim (1/J_0)\exp{(l/\zeta)}$.

The time evolution of $O_{\beta}(t)$ thus starts with it essentially time independent at early times, as discussed above.  Around $t\sim (1/J_0)\exp{(r/\zeta)}$ all the eigenstates that differ in between sites $r_1$ and $r_2$ begin to dephase.  The number of such states is $\sim\exp{(sr)}$ where $s$ is the entropy per unit length which depends on $\beta$ and decreases with increasing $|\beta|$.  The time-dependent part of $O_{\beta}(t)$ then becomes a sum of an effectively random contribution from each dephased eigenstate, so its magnitude rapidly decays around this time by a factor of $\sim\exp{(sr/2)}$.  At later times, the distance over which other l-bits cause dephasing is $l\sim\zeta\log{(J_0t)}$.  There are two cases: if either $r_1$ or $r_2$ is at the end of a semi-infinite chain, then the number of dephased eigenstates grows as $\sim\exp{(sl)}$, since the extra dephasing can only come from one side at the end of such a semi-infinite system.  This results in the time-dependent part of the OTO correlator oscillating quasiperiodically inside a power-law decaying envelope so that
\begin{equation}
|O_{\beta}(t)-O_{\beta}(t\rightarrow\infty)|^2\sim (J_0t)^{-s\zeta}
\label{oto_decay}
\end{equation}
at the end of the chain.  Note that this behavior does not depend on $r=|r_1-r_2|$ in this time regime.  If, on the other hand, we are in the bulk of an infinite chain then there is more dephasing, since other l-bits on both sides can contribute to the dephasing.  In this case the number of dephased eigenstates grows as $\sim\exp{(s(2l-r))}$ so there is more dephasing in this late time range as compared to the end of the chain.  This results in instead
\begin{equation}
|O_{\beta}(t)-O_{\beta}(t\rightarrow\infty)|^2\sim (J_0t)^{-2s\zeta}\exp{(sr)}
\end{equation}
in the bulk of an infinite chain in this late-time regime.

If we simply do the disorder average $\overline{O_{\beta}(t)}$ then these power laws due to dephasing are not seen, since they oscillate incoherently between samples and their disorder average is thus zero.  To see the power laws the
disorder average that we take is $\overline{|O(t)|^2}$.
For two generic local operators  $\hat{W}(r_1,0)$ and $\hat{V}(r_2,0)$, this disorder average in an infinite system
will exhibit a late time power law decay to a constant,
\begin{equation}
\overline{|O_{\beta}(t)|^2}\sim \frac{A}{t^{\alpha}}+\gamma ~,
\label{OTO_scaling}
\end{equation}
where $\gamma\geq 0$ is determined by the overlap of $\hat{W}(r_1,0)$ and $\hat{V}(r_2,0)$ with the local integrals of motion $\hat{\tau}^z$, $A\geq 0$ is determined by their overlaps with the $\hat\tau^{x,y}$'s, and the exponent $\alpha$ is as given above.


Finally, we would like to briefly discuss the OTO correlator for the noninteracting Anderson localized phase. The effective model for an Anderson localized phase is \eqref{eff_H} with all the interaction terms set to zero.  Therefore, the only OTO correlators of the $\tau$'s that are time-dependent are at the same site in \eqref{eff_H}.  If we look at the OTO correlators of the bare operators at two different sites, time-dependence arises only to the extent that single-particle eigenstates have weight at both of those sites.  The frequencies that appear in the time-dependence are only the single-particle energies and there is no dephasing, since there are no interactions.  So $O(t)$ for a given sample remains quasiperiodic and does not decay to a constant at long time, even after thermal averaging.


\section{Numerical simulation}
\label{sec:numeric}

To test the above arguments, we begin by numerically studying the OTO correlator in the effective model and then study three more ``realistic'' spin chain models. Throughout this section, we use open boundary conditions and put the $\hat W(r_1,0)$ on the left boundary, so that it will only propagate in one direction. Unless specified otherwise, the disorder average in the calculation is over $10^4$ samples, such that the error bar is almost invisible in our plots in most regimes.

\subsection{The effective model}
\label{sec:num_eff}

\begin{figure}[hbt]
\centering
\vspace*{-0cm}
\includegraphics[scale=.4]{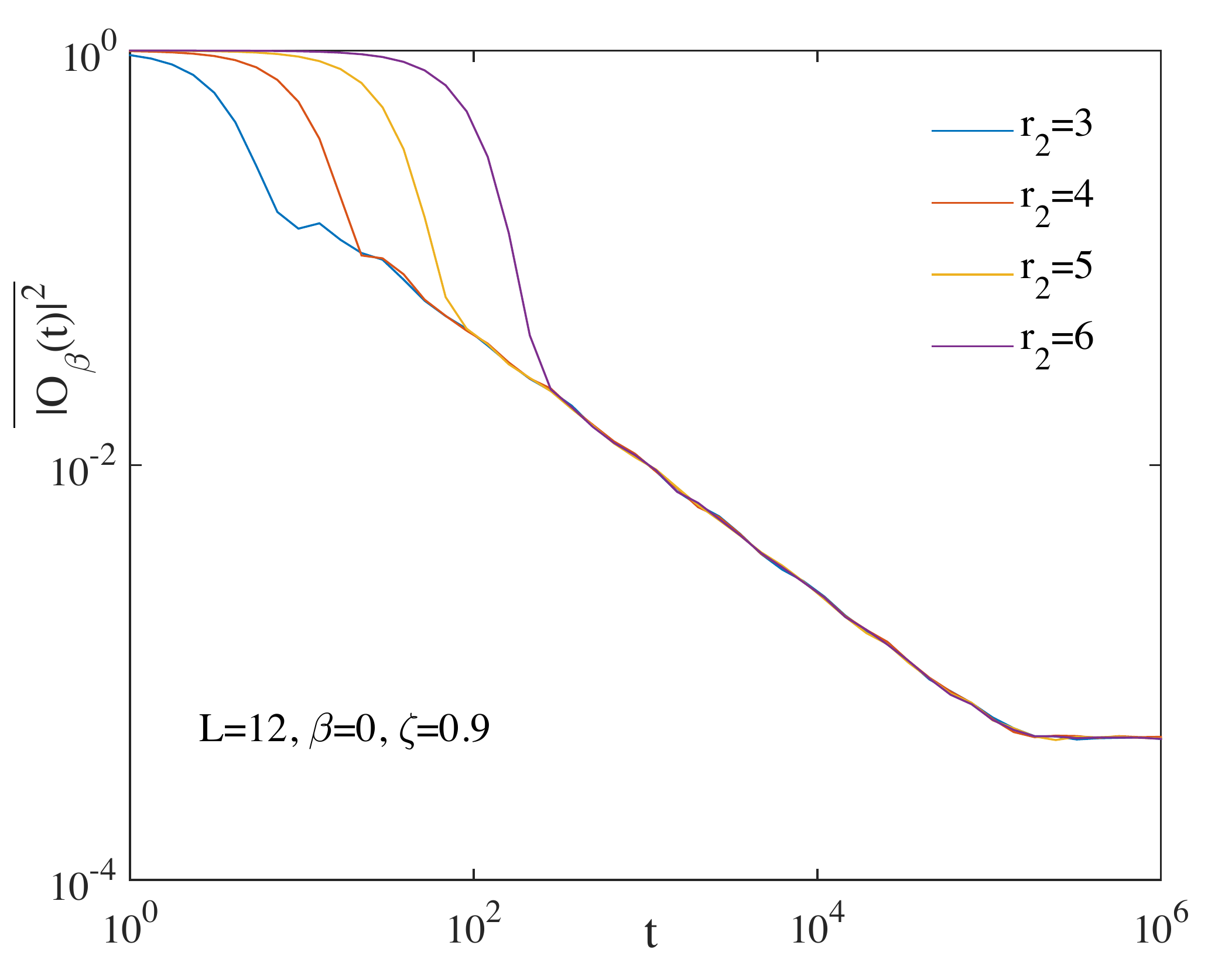}
\caption{The disorder average of the square of the thermally averaged OTO correlator $\overline{|O_\beta(t)|^2}$ in the effective model at $\beta=0$ on a log-log plot, where $\hat{W}(r_1,0)=\hat{\tau}_{r_1}^x$ and $\hat{V}(r_2,0)=\hat{\tau}_{r_2}^x$ with $r_1=1$ and $r=|r_2-r_1|$. The decay length is $\zeta=0.9$ in units of the lattice spacing.}
\label{fig:OTO_eff_beta_0}
\end{figure}

\begin{figure}[hbt]
\centering
\vspace*{-0cm}
\includegraphics[scale=.4]{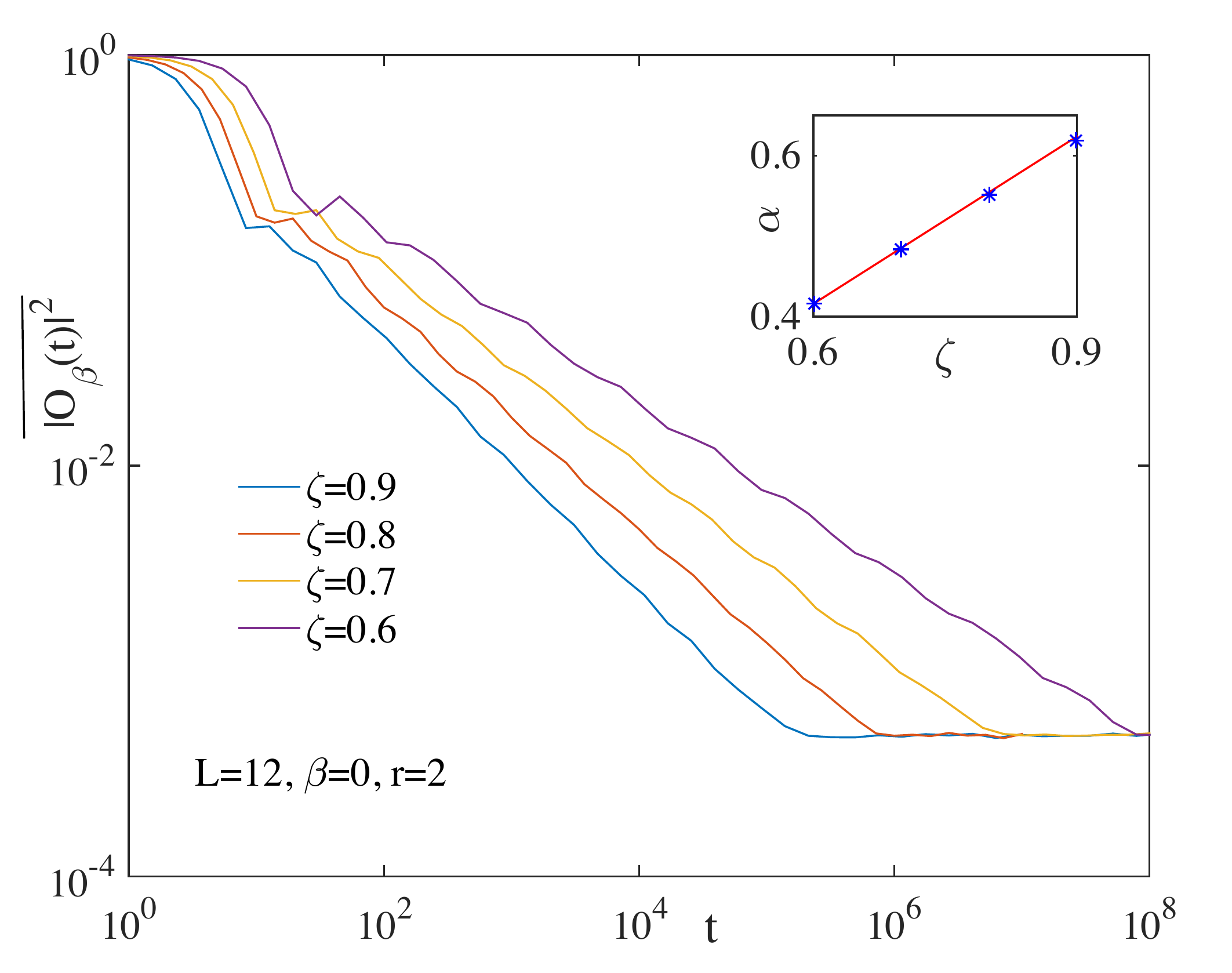}
\caption{$\overline{|O_\beta(t)|^2}$ in the effective model with different decay lengths $\zeta$ on a log-log plot. Here we choose $\hat{W}(r_1,0)=\hat{\tau}_{1}^x$ and $\hat{V}(r_2,0)=\hat{\tau}_{3}^x$.  The power law exponents $\alpha$ (blue points) are shown in the inset and they match up with the theoretical prediction (red curve) from Eq. \ref{oto_decay}.}
\label{fig:OTO_eff_xi_com}
\end{figure}

In this section, we check numerically the OTO correlators in the effective model defined in Eq. \eqref{eff_H} with $\hat{W}(r_1,0)=\hat \tau_{r_1}^{x}$ and $\hat{V}(r_2,0)=\hat \tau_{r_2}^x$.  We take the system size to be $L = 12$ and calculate the disorder average of the square of the OTO correlator (denoted by $\overline{ |O_\beta(t)|^2}$) for different separations (Fig. \ref{fig:OTO_eff_beta_0}) and decay lengths (Fig. \ref{fig:OTO_eff_xi_com}) at $\beta=0$.  In the calculation, $J_0$ is fixed to be of order 1, and the coupling strength for each interaction term in Eq. \eqref{eff_H} is random and taken from a uniform distribution in the interval $[-2^{-l/2}e^{-l/\zeta}, 2^{-l/2}e^{-l/\zeta}]$, where $l$ is the distance between the pair of l-bits in that interaction term that are farthest apart.  The factors of $2^{-l/2}$ are there so that when all contributions are added up the resulting $\hat{J}^{eff}$ decays exponentially with decay length $\zeta$.  This is only an approximation to what the effective model would be for a real spin chain.  More realistically, the magnitudes of the interactions should more broadly distributed (something like log-normal) and they should be highly correlated since they all arise from the same instance of a local Hamiltonian with only of order $L$ random parameters.  We do not vary $\beta$ for this effective model, since our approximation ignores the detailed correlations among the interactions that give the proper dependence of $\zeta$ on $\beta$.

In Fig. \ref{fig:OTO_eff_beta_0}, we show $\overline{|O_\beta(t)|^2}$ for $\hat\tau^x$ for separations $r=|r_2-r_1|$ ranging from 2 to 5.  As expected, at early times the OTO correlator is nearly constant, since we are outside of the ``logarithmic light cone'' and the operator $\hat W(r_1,t)$ has not yet spread enough to have any substantial effect at $r_2$.\cite{Bardarson2012, Vosk2013, Serbyn2013a, Serbyn2014a, Huse2014, Kim2014}  The logarithmic growth of local operators with time is in contrast with the ballistic growth in chaotic nonrandom systems.\cite{Shenker2014} It is a manifestation of the slow dynamics in MBL. When the time finally passes through the edge of the ``light cone'' strong dephasing starts and the OTO correlators decrease by a factor that is exponential in $r$.  After this we see the expected power law regime with the OTO correlator independent of $r$, since $r_1$ is at the end of the chain.  At the very latest times all eigenstates of this finite system have fully dephased and this correlator saturates to a small value $\sim 2^{-(L-1)}$ set by the sample size.

In Fig. \ref{fig:OTO_eff_xi_com}, we fix the separation to $r=2$ and vary the decay length. Obviously, the OTO correlator decays faster as $\zeta$ increases and thus produces more dephasing. The exponent of the power law decay depends on
$\zeta$ as shown in the inset of Fig. \ref{fig:OTO_eff_xi_com}, in agreement with Eq. \ref{oto_decay}.  The regular ``wiggles'' in the decay, which are fairly apparent for $\zeta=0.6$, are an artifact of our approximation to the effective model, which gives all samples the same distribution of interactions at each distance.  This feature also makes the crossover from the power-law decay to the late-time finite-size saturation sharper than it should be in both Figs. \ref{fig:OTO_eff_beta_0} and \ref{fig:OTO_eff_xi_com}.  



\begin{figure}[hbt]
\centering
\vspace*{0 cm}
\includegraphics[scale=.4]{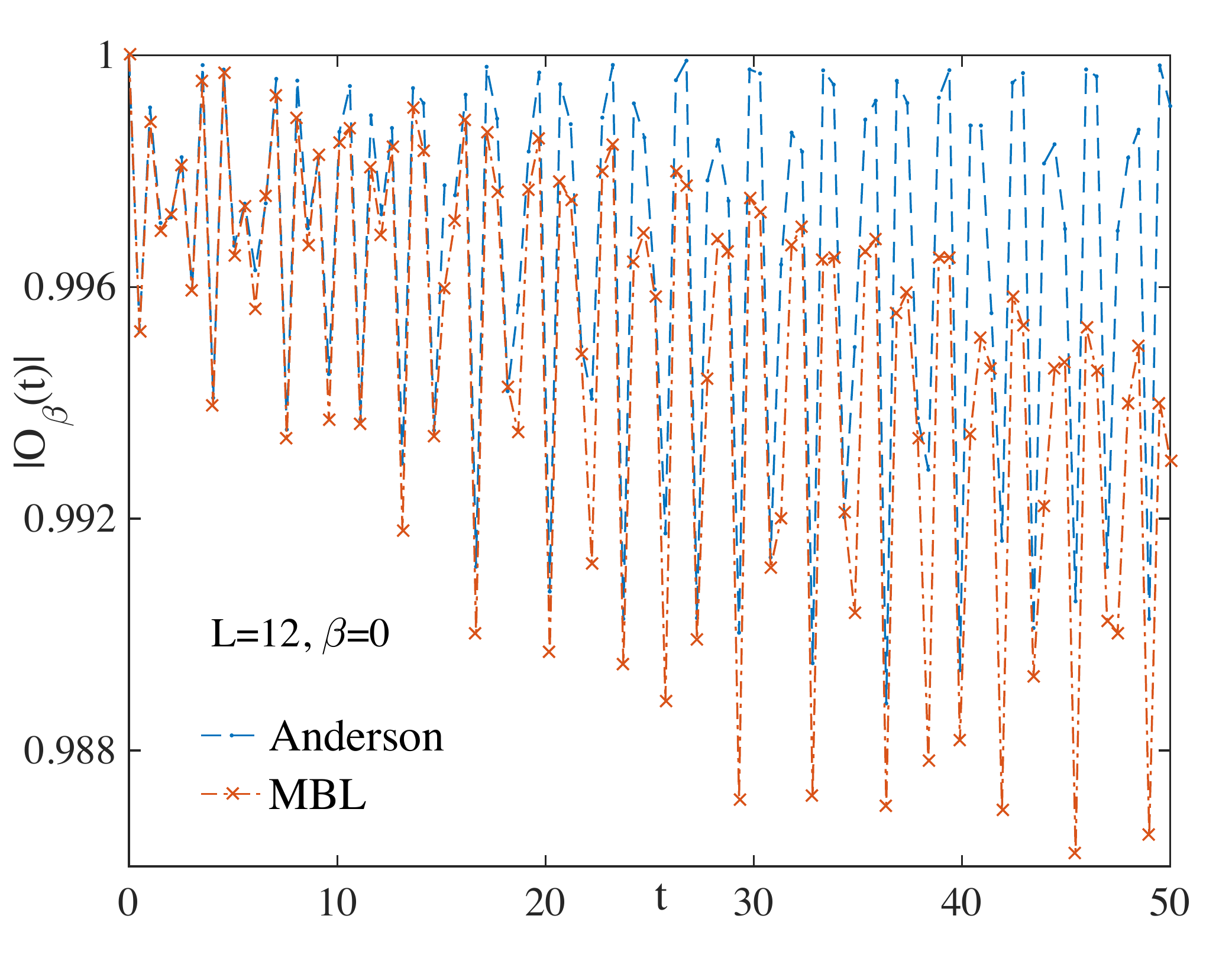}
\caption{$|O_\beta(t)|$ at early times for a noninteracting Anderson localized system and for the MBL phase.  These are for only one realization of the random fields and we use the same realization for both cases.  $\hat W(t=0)=\hat\sigma^z_1$ and $\hat V(t=0)=\hat\sigma^z_3$ with the Hamiltonian as defined in Eq. \eqref{trans_field}.  The blue points (connected with dashed lines) are for the Anderson localized case with $W=10$ and $h_z=0$. The orange crosses are for the MBL phase with $W=10$ and $h_z=0.1$.}
\label{fig:Ising_early}
\end{figure}

\subsection{Ising spin chain with uniform longitudinal field and random transverse field}
\label{sec:trans}
In this section, we numerically study an Ising spin chain with a uniform longitudinal field and a random or uniform transverse field. 
We use the exact diagonalization method to diagonalize its Hamiltonian and then calculate the OTO correlator directly. The Hamiltonian is
\begin{equation}
\hat H=-\sum_i\hat\sigma^z_i\hat\sigma^z_{i+1}-\sum_i h_i\hat\sigma^x_i-h_z\sum_i \hat\sigma^z_i ~.
\label{trans_field}
\end{equation}
When the transverse field is random, we draw each $h_i$ from a uniform distribution on the interval $[-W,W]$.
In the absence of the longitudinal field $h_z$, this is the transverse field Ising model and can be mapped to a noninteracting fermion model after a Jordan-Wigner transformation.  If the transverse field is random, this noninteracting model is Anderson localized.  If we take both $\hat W(t=0)$ and $\hat V(t=0)$ to be $\hat\sigma^z$ operators, when the disorder strength $W\gg 1$, the OTO correlator $|O_{\beta}(t)|$ will remain close to one for $h_z=0$.  As we discussed above, if $\hat W(t=0)$ and $\hat V(t=0)$ have overlap with the same l-bit, the OTO correlator oscillates quasiperiodically with time and does not converge to a constant. This is shown in Fig. \ref{fig:Ising_early}, where the blue curve oscillates around 0.995. The frequencies are determined by the single particle energies and are of order one.

\begin{figure}[hbt]
\centering
\vspace*{0 cm}
\includegraphics[scale=.4]{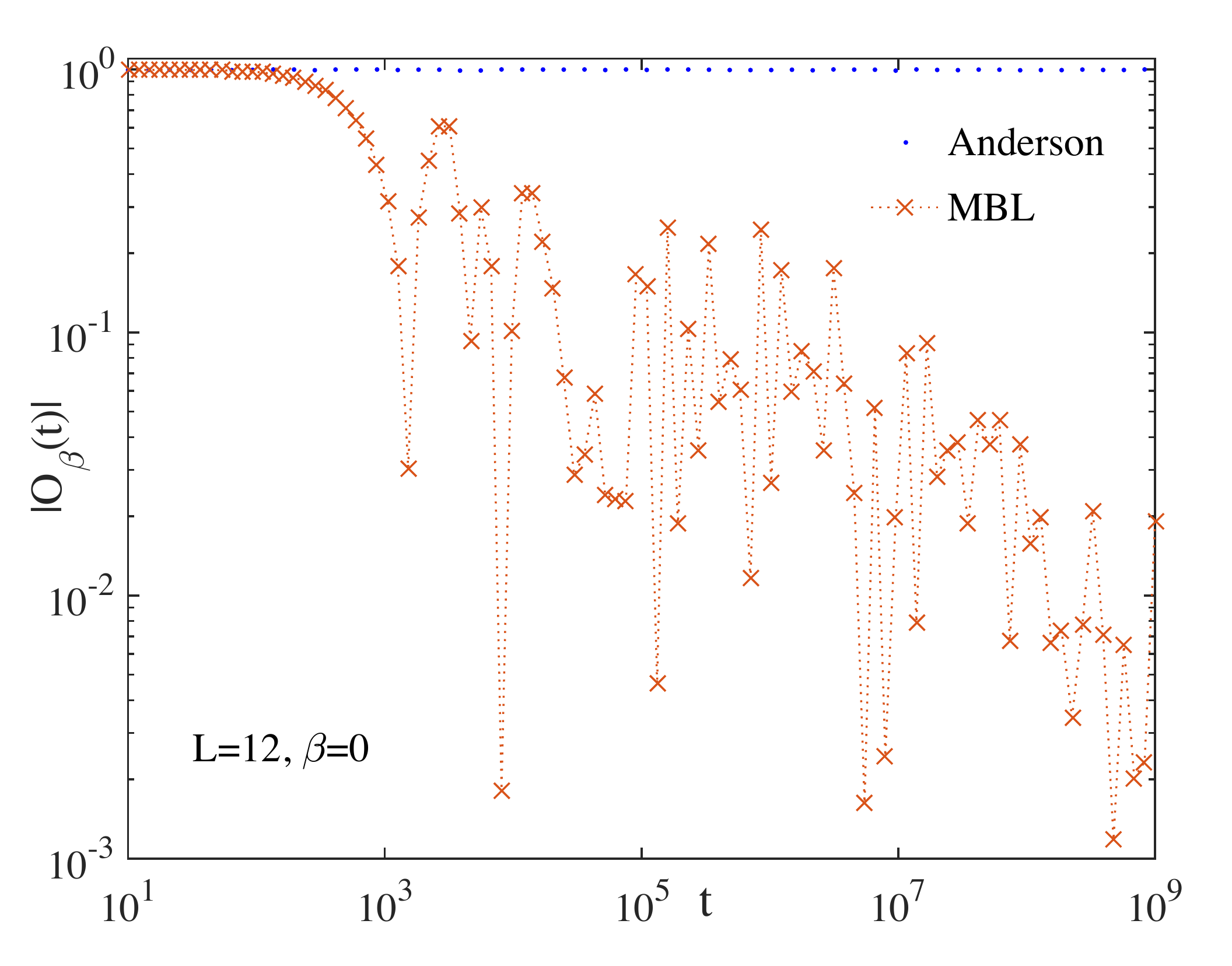}
\caption{The same quantities in the same samples as shown in Fig. \ref{fig:Ising_early}, but here shown to long time on log scales.}
\label{fig:Ising_late}
\end{figure}

The behavior of the Anderson localized case $h_z=0$ is unstable to dephasing when interactions $h_z\neq 0$ are added. In the model defined in Eq. \eqref{trans_field}, if the disorder strength is strong, a small $h_z$ puts the system in the MBL phase. In Fig. \ref{fig:Ising_early}, we compare single samples with precisely the same random transverse fields with and without an interaction ($h_z=0$ and $h_z=0.1$).  At the earliest times both systems show roughly the same pattern of oscillation.  The onset of the dephasing due to the interaction becomes quite apparent in this plot around $t=25$.  As shown in Fig. \ref{fig:Ising_late}, this damping due to dephasing continues to very long time, with the MBL system's OTO correlator oscillating within a power-law decaying envelope.


\begin{figure}[hbt]
\centering
\vspace*{-0cm}
\includegraphics[scale=.7]{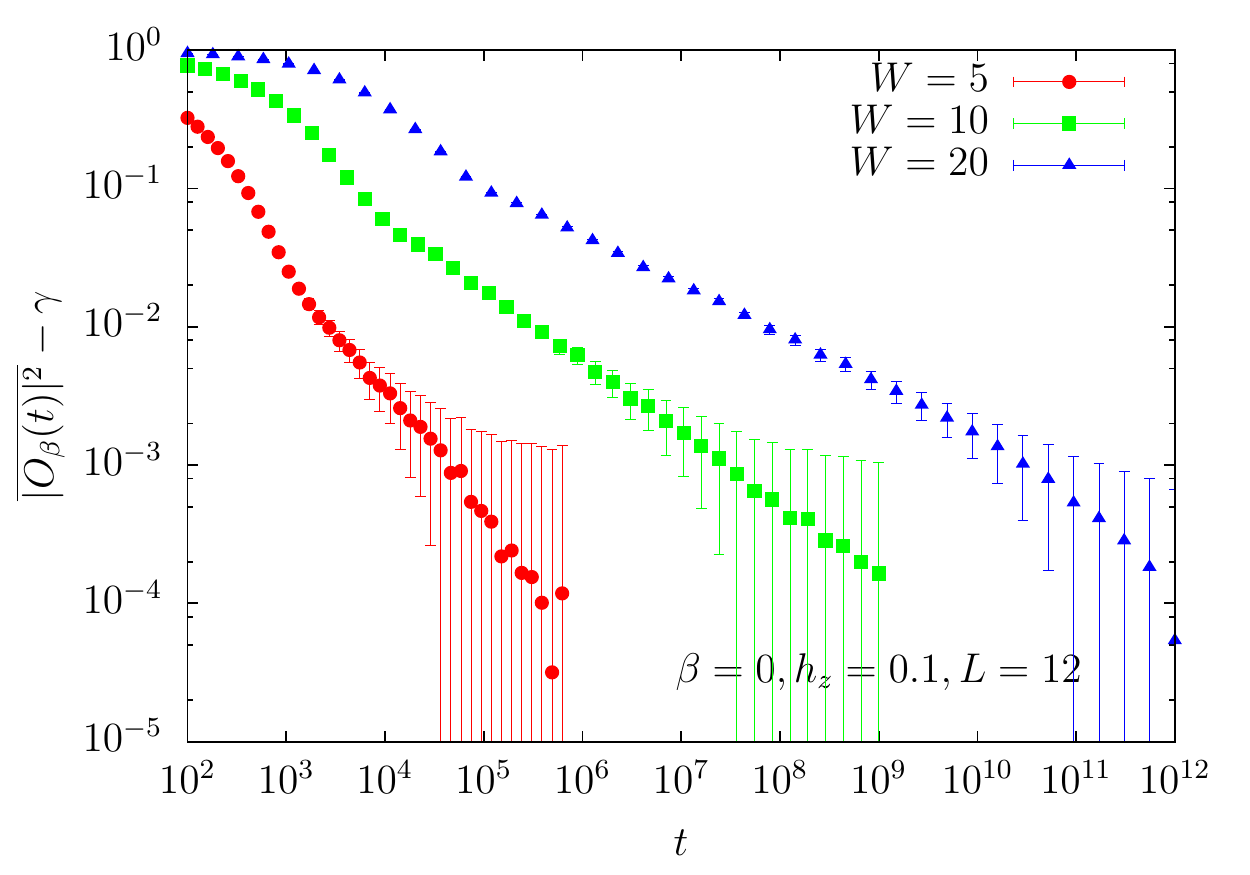}
\caption{$\overline{|O_\beta(t)|^2}-\gamma$ with $\hat{W}(r_1,0)=\hat{\sigma}_{1}^z$ and $\hat{V}(r_2,0)=\hat{\sigma}_{3}^z$ at various $W$ in a log-log plot. The Hamiltonian is the Ising model in a random transverse field and a uniform longitudinal field defined in Eq. \eqref{trans_field}. The parameters are $L=12$, $h_z=0.1$ and $\beta=0$.  The standard deviation for $\overline{|O_\beta(t)|^2}$ is always less than $1\%$. The growth of the error bars with  time on the plot is coming from two aspects: (1) Once we subtract $\gamma$, the relative error for $\overline{|O_\beta(t)|^2}-\gamma$ increases with the time and looks large for the last several points. (2) The log-log scale exaggerates the error bars at late times. $\gamma$ is evaluated by taking disorder average on $10^4$ samples at around $t=10^{18}$ with relative error  not larger than $1\%$.}
\label{fig:OTO_W}
\end{figure}

\begin{figure}[hbt]
\centering
\vspace*{-0cm}
\includegraphics[scale=.7]{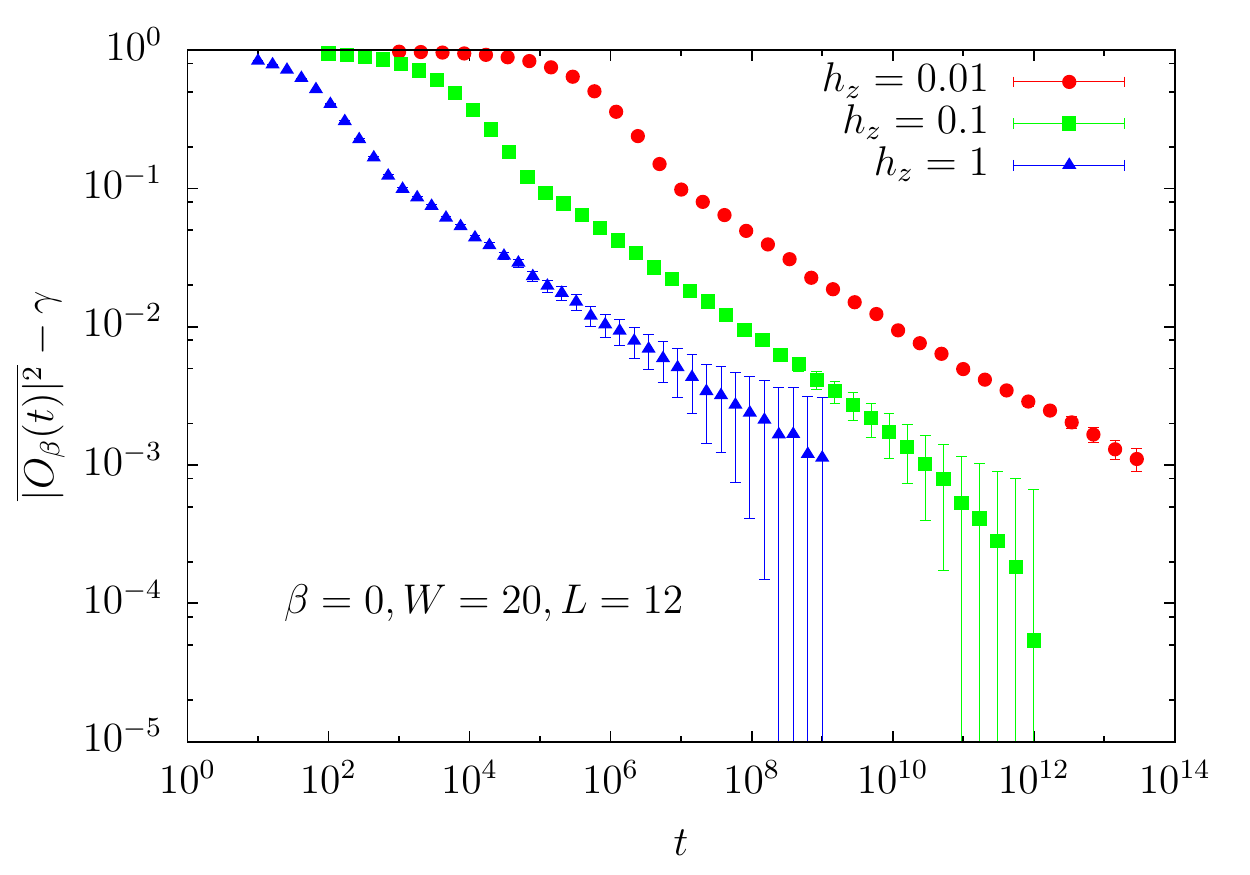}
\caption{$\overline{|O_\beta(t)|^2}-\gamma$ with $\hat{W}(r_1,0)=\hat{\sigma}_{1}^z$ and $\hat{V}(r_2,0)=\hat{\sigma}_{3}^z$ at various $h_z$ in a log-log plot. The Hamiltonian is the Ising model in a random transverse field and a uniform longitudinal field defined in Eq. \eqref{trans_field}. The parameters are $L=12$, $W=20$ and $\beta=0$.}
\label{fig:OTO_hz}
\end{figure}

\begin{figure}[hbt]
\centering
\vspace*{-0 cm}
\includegraphics[scale=.4]{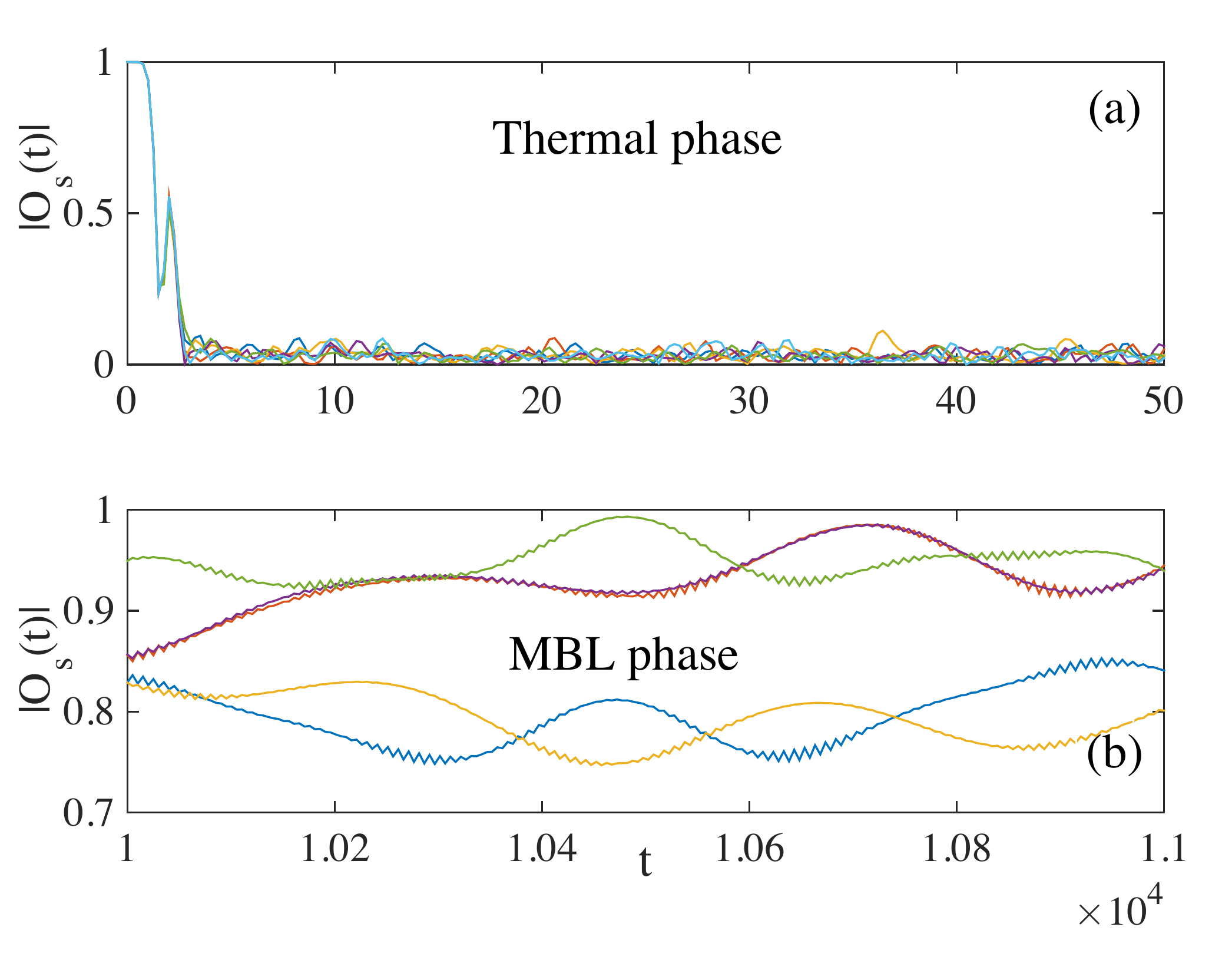}
\caption{$|O_s(t)|$ for five eigenstates in the middle of the energy spectrum for the model in Eq. \eqref{trans_field}, where $\hat{W}(r_1,0)=\hat{\sigma}_{1}^z$ and $\hat{V}(r_2,0)=\hat{\sigma}_{3}^z$. The system size is $L=10$. (a) is the thermal phase with $h=1.05$ and $h_z=0.5$.\cite{Banuls2011} (b) is the MBL phase with $W=10$ and $h_z=0.1$.}
\label{fig:OTO_all_trans}
\end{figure}


For each disorder realization, $|O_\beta(t)|$ has damped pseudorandom oscillations within a power-law decaying envelope function. 
Once we take the disorder average, these oscillations are fully dephased and only the power-law decay behavior (\ref{OTO_scaling}) is left. In an infinitely long system, $|O_{\beta}(t)|^2$ will eventually saturate to a constant $\gamma$ determined by the overlap between the initial operators and l-bits. In our numerical calculation for small $L$, to suppress the finite size effect, $\gamma$ is estimated by choosing a constant slightly smaller than the long time limit of $|O_{\beta}(t)|^2$ (within the statistic error bar).  In Fig. \ref{fig:OTO_W} and Fig. \ref{fig:OTO_hz}, we plot $\overline{|O_\beta(t)|^2}-\gamma$ on logarithmic scales for different $W$ and $h_z$. As long as $h_z$ is nonzero, $\overline{|O_\beta(t)|^2}-\gamma$ decays as a power-law in the long time regime. The slope decreases as $W$  increases, since the decay length $\zeta$ decreases as the system gets more strongly localized.   We find that the slope, and thus the decay length $\zeta$, is not sensitive to $h_z$, suggesting that here the decay length is set by the localization length at $h_z=0$.  The value of $h_z$ does affect the time at which the dephasing starts.
The power-law decay is caused by destructive interference between different eigenstates. We verify this point by further studying the OTO correlator for single eigenstates.  Fig. \ref{fig:OTO_all_trans}(b) shows the $|O_s(t)|$ for some excited eigenstates in the MBL phase. They oscillate at different frequencies and never decay to zero,  in contrast with the thermal phase shown in Fig. \ref{fig:OTO_all_trans}(a), where $|O_s(t)|$ for the highly excited eigenstates decay quickly to near zero. 

\subsection{spin-$1/2$ Heisenberg model in a random field}
\label{sec:hei}
To show that the long-time power-law decay of the OTO correlator is a generic feature of the MBL phase, we consider another disordered spin chain model which has a MBL phase. This model is a $S=1/2$ Heisenberg chain in a random magnetic field, governed by the Hamiltonian
\begin{equation}
\hat H=\sum_i\left(\hat{\sigma}^x_i\hat{\sigma}^x_{i+1}+\hat{\sigma}^y_i\hat{\sigma}^y_{i+1}+\hat{\sigma}^z_i\hat{\sigma}^z_{i+1}\right)+\sum_i h_i\hat\sigma^z_i ~,
\label{hei_mod}
\end{equation}
where the random field $h_i$ takes a uniform distribution between $[-W,W]$. This model has been used extensively to study the MBL phase and the associated localization-delocalization phase transition.\cite{Pal2010, Luca2013, Bauer2013, Nanduri2014, Luitz2015} Here we take the system size to be $L=12$ and calculate the OTO correlator for different $\beta$ and different $W$.


\begin{figure}[hbt]
\centering
\vspace*{-0 cm}
\includegraphics[scale=.7]{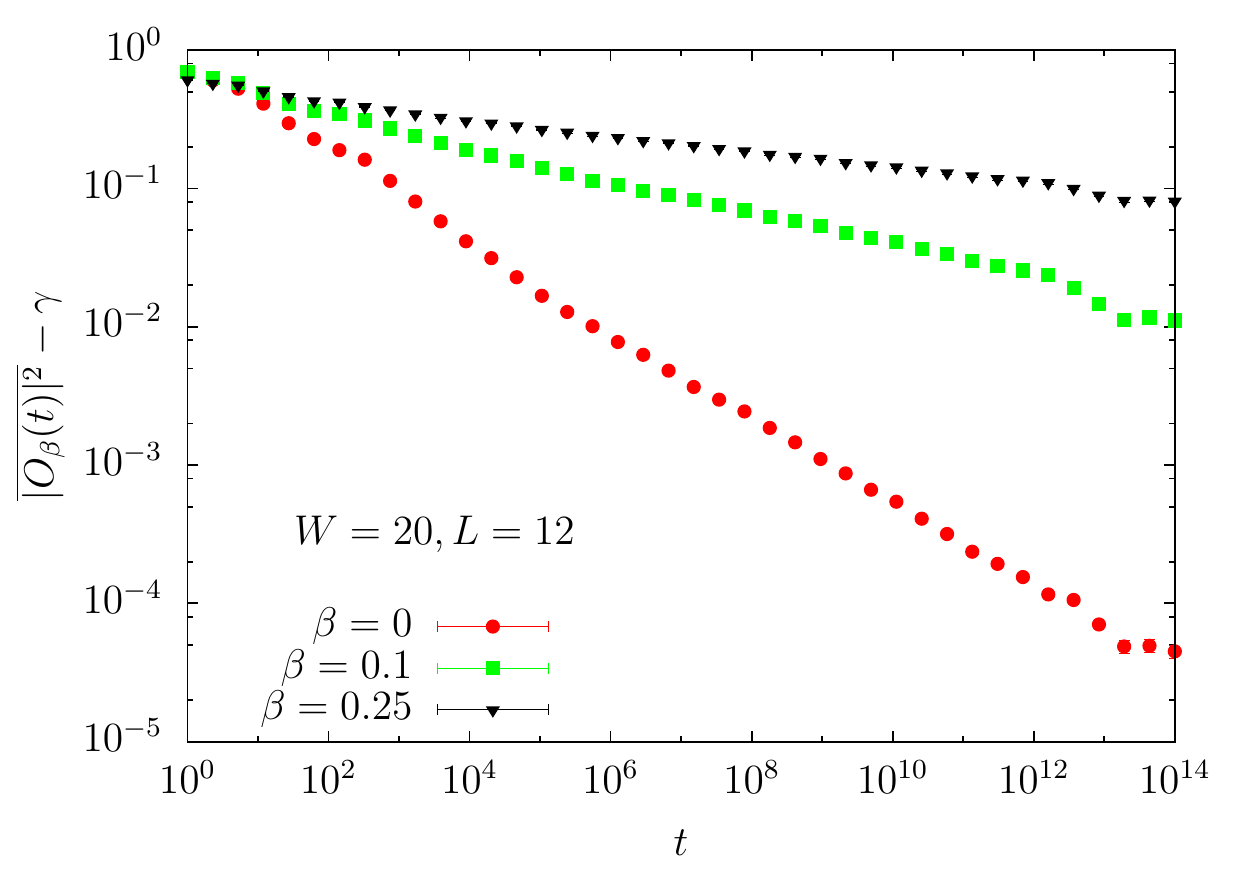}
\caption{$\overline{|O_\beta(t)|^2}-\gamma$ with $\hat{W}(r_1,0)=\hat{\sigma}_{1}^x$ and $\hat{V}(r_2,0)=\hat{\sigma}_{3}^x$ for the Heisenberg model in a random field at various $\beta$ on a log-log plot. The system parameters are $L=12$ and the random field $h_i\in [-20,20]$.}
\label{fig:Hei_beta_W20}
\end{figure}

In Fig. \ref{fig:Hei_beta_W20}, we plot $\overline{|O_\beta(t)|^2}-\gamma$ inside the MBL phase on a log-log plot at various $\beta$ for $W=20$.  The expected power law behavior is seen.  As $\beta$ increases from $0$ to $0.25$, the slope becomes shallower.  As we discussed in Eq. \eqref{oto_decay}, the power law exponent $\alpha$ depends on both the entropy density $s$ and the decay length $\zeta$.  The entropy per spin $s$ decreases with increasing $\beta$ and is equal to $\log(2)$, $0.38$ and $0.16$ for the three values of $\beta$ shown in Fig. \ref{fig:Hei_beta_W20}.  Since $\zeta=\alpha/s$, we find that the values of $\zeta$ are $0.57$, $0.43$ and $0.32$, showing that the decay length $\zeta$ does decrease significantly as $\beta$ increases: the system, as expected, becomes more localized as the temperature is reduced.

\begin{figure}[hbt]
\centering
\vspace*{-0 cm}
\includegraphics[scale=.7]{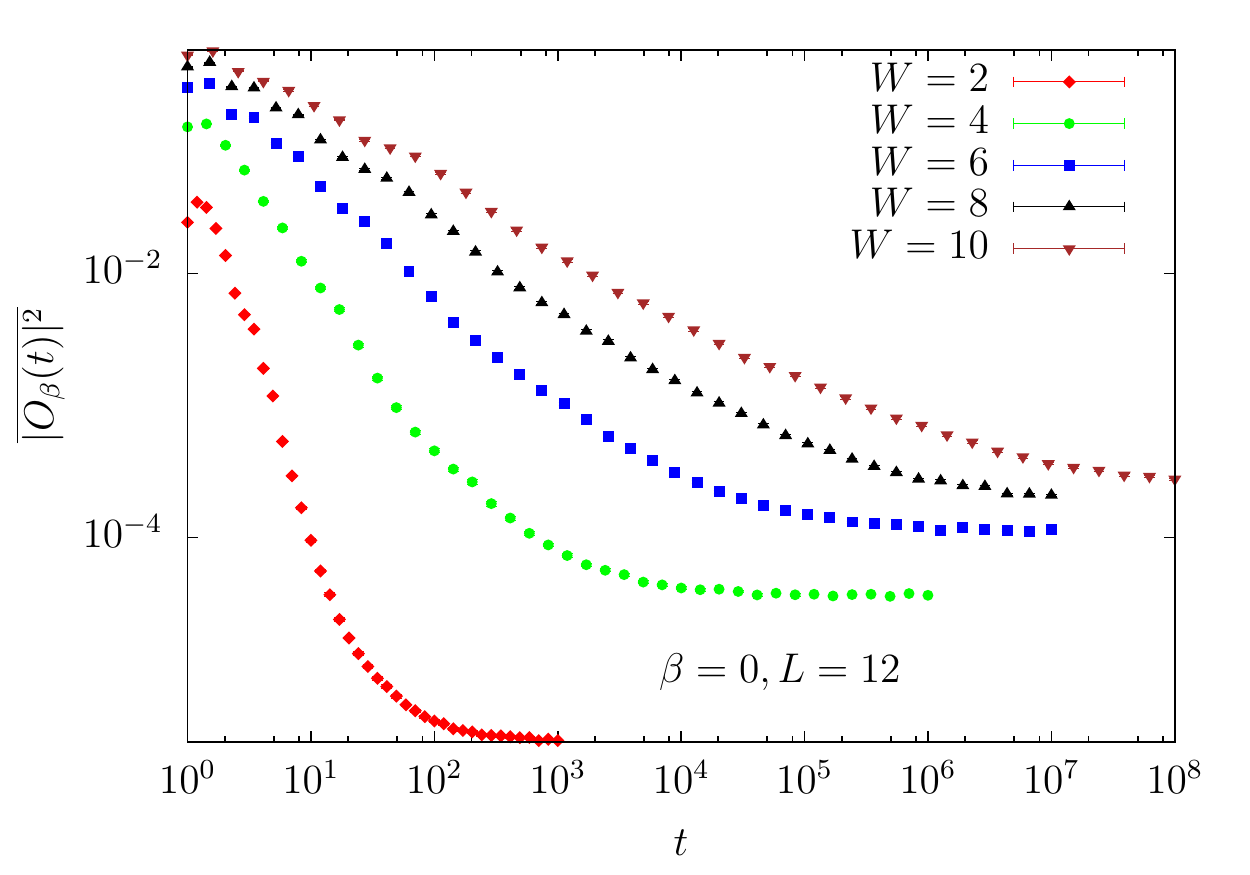}
\caption{$\overline{|O_\beta(t)|^2}$ with $\hat{W}(r_1,0)=\hat{\sigma}_{1}^x$ and $\hat{V}(r_2,0)=\hat{\sigma}_{3}^x$ for the Heisenberg model with a random field at $\beta=0$ and various $W$ on a log-log plot.}
\label{fig:Hei_h}
\end{figure}

\begin{figure}[hbt]
\centering
\vspace*{-0 cm}
\includegraphics[scale=.7]{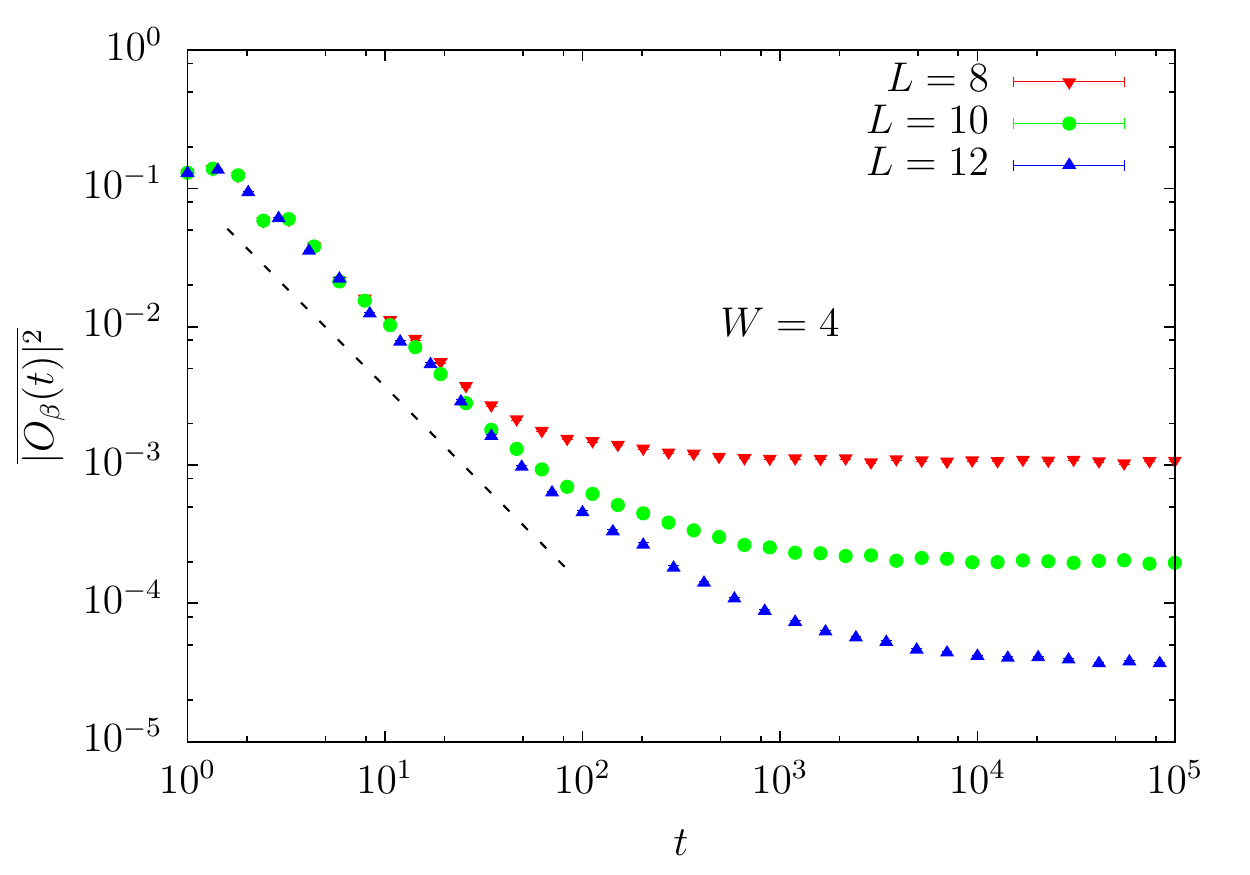}
\caption{$\overline{|O_\beta(t)|^2}$ with $\hat{W}(r_1,0)=\hat{\sigma}_{1}^x$ and $\hat{V}(r_2,0)=\hat{\sigma}_{3}^x$ for the Heisenberg model in a random field with various system sizes. The  random field $h_i\in [-4,4]$ and $\beta=0$.}
\label{fig:Hei_beta_W4}
\end{figure}

For this model \eqref{hei_mod} we have also examined the behavior of the OTO correlator near the phase transition (which is thought to occur somewhere between $W=7$ and $W=10$ \cite{Luitz2015}) and in the thermal phase, as shown in Figs. \ref{fig:Hei_h} and \ref{fig:Hei_beta_W4}. Notice that the curves are for $\overline{|O_{\beta}(t)|^2}$ without subtracting a long-time constant. We find that even fairly deep in the thermal phase, the decay is still close to power law before it slows down and saturates to a small constant. The final saturation is a finite-size effect. In Fig. \ref{fig:Hei_beta_W4}, we draw $\overline{|O_{\beta}(t)|^2}$ for various system sizes at $W=4$. The power law behavior is quite obvious before the finite-size effect shows up.  This power law decay seems to be the case for all the different local Pauli operators that we tried for $\hat{V}$ and $\hat{W}$.  It appears that the OTO is always coupling somehow to the slow transport of the conserved quantities (energy and total $\sigma^z$), perhaps in a way that also involves rare region Griffiths effects \cite{gopa}.  We now turn to a Floquet spin chain to see if this feature changes in the absence of conservation laws. 

\subsection{Floquet spin-$1/2$ chain}
\label{sec:flo}
In this section, we study a disordered system subject to periodic driving.  This Floquet system has no conserved densities.  Previous studies show that disordered Floquet systems can have both a thermal phase and a MBL phase.\cite{Ponte2015, Lazarides2015}  It has been shown that Floquet systems without conservation laws can thermalize faster and more completely than the corresponding Hamiltonian systems which necessarily conserve at least energy.\cite{Zhang2015, Zhang2016}  Here we numerically calculate the OTO correlator in both phases of such a Floquet model and compare the results with other models.  The properties of this periodically driven system are determined by the unitary time evolution operator over one period, i.e., the Floquet operator. Following Ref.\ \onlinecite{Zhang2016}, we consider this Floquet operator:
\begin{align}
\hat{U}_F=\exp\left[-i\frac{\tau}{2}\hat{H}_x\right]\exp[-i\tau \hat{H}_z]\exp\left[-i\frac{\tau}{2}\hat H_x\right] ~,
\label{flo_op}
\end{align}
where
\begin{align}
&\hat{H}_x=\sum_{j=1}^{L}g\Gamma\hat\sigma_j^x\nonumber\\
&\hat{H}_z=\sum_{j=1}^{L-1}\hat\sigma_j^z\hat\sigma_{j+1}^z+\sum_{j=1}^L(h+g\sqrt{1-\Gamma^2}G_j)\hat\sigma_j^z ~.
\end{align}
This model is a periodically driven system with period $2\tau$. $\hat{U}_F$ is chosen in this particular way so that it enjoys the time reversal symmetry. The eigenstates are real in the $\{\sigma^z\}$ basis, which simplifies the diagonalization of the $\hat{U}_F$ operator.  In the numerical calculation, we choose open boundary condition with $L=12$. The system parameters are $(g,h,\tau)=(0.9045, 0.8090, 0.8)$, and for each sample $\{G_j\}_{j=1}^L$ is a set of independent Gaussian standard normal random variables.\cite{Zhang2016}

$\Gamma$ here controls the transverse field strength in $\hat{H}_x$ and also the disorder strength in the longitudinal field in $\hat{H}_z$. As $\Gamma\to 1$, the disorder strength goes to zero and this model thermalizes rapidly.\cite{Zhang2015,Kim_H2014} As $\Gamma\to 0$, the transverse field in $\hat{H}_x$ goes to zero and it is deep inside MBL phase with the $\{\sigma^z\}$ as the local integrals of motion. Ref.\ \onlinecite{Zhang2016} shows that the phase transition occurs at around $\Gamma=0.3$.

\begin{figure}[hbt]
\centering
\vspace*{-0 cm}
\includegraphics[scale=.7]{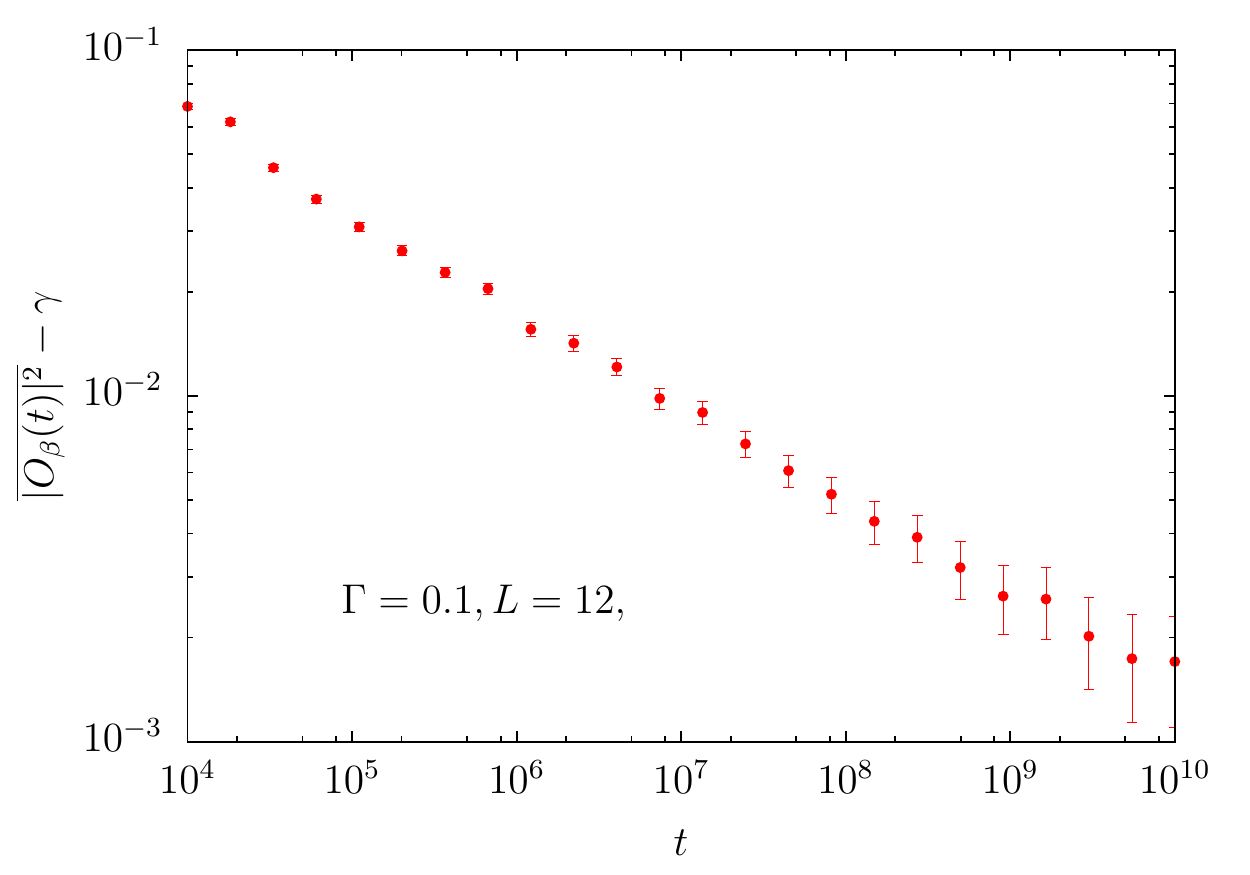}
\caption{$\overline{|O_\beta(t)|^2}-\gamma$ with $\hat{W}(r_1,0)=\hat{\sigma}_{1}^x$ and $\hat{V}(r_2,0)=\hat{\sigma}_{3}^x$ at $\Gamma=0.1$ on the log-log plot. This Floquet spin model is defined in Eq. \eqref{flo_op} with system size $L=12$. }
\label{fig:Flo_MBL}
\end{figure}


\begin{figure}[hbt]
\centering
\vspace*{-0 cm}
\includegraphics[scale=.7]{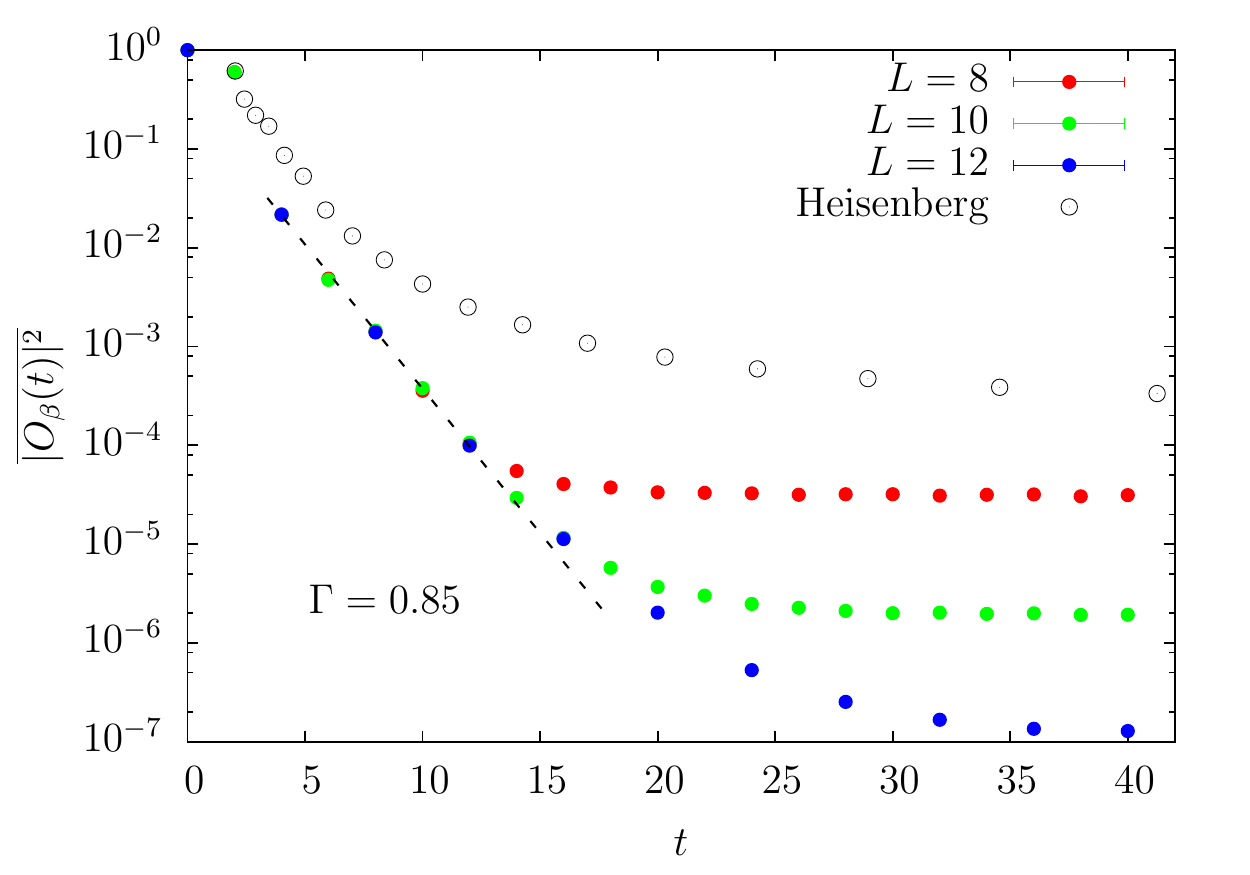}
\caption{$\overline{|O_{\beta}(t)|^2}$ with $\hat{W}(r_1,0)=\hat{\sigma}_{1}^x$ and $\hat{V}(r_2,0)=\hat{\sigma}_{3}^x$ on a semi-log plot. The Floquet spin model is defined in Eq. \eqref{flo_op} with system size $L=12$ and $\Gamma=0.85$. The black empty circles are for the thermal phase of Heisenberg model in a random field with $W=2$ and $L=12$ (we shift this curve upward for better comparison).}
\label{fig:Flo_thermal2}
\end{figure}

For this Floquet model, the thermal average is uniform over all states since there are no conservation laws. This corresponds to taking $\beta=0$ in the OTO correlator. We first study the OTO correlator in the MBL phase and  find that the OTO correlator at $\Gamma=0.1$ has power law decay as expected (Fig. \ref{fig:Flo_MBL}).  We also examine the thermal phase with $\Gamma>0.3$. As shown in Fig. \ref{fig:Flo_thermal2}, the decay of the time-dependent part of the OTO correlator fits well to a simple exponential function of time when we are deep in the thermal phase of this model:
\begin{align}
\overline{|O_{\beta}(t)|^2}\sim  A\exp(-\lambda t) ~,
\end{align}
where $\lambda$ increases as we move deeper into the thermal phase. In the numerical calculation, the finite-size effect will slow down the exponential decay and $\overline{|O_{\beta}(t)|^2}$ will eventually saturate to a small constant.  Nevertheless, the exponential decay is obvious and is much faster than the power-law decay in the thermal phase of the time-independent Hamiltonian studied in the previous section (black empty circles in Fig. \ref{fig:Flo_thermal2}).  This exponential decay behavior has also been observed in the holographic conformal field theory which has  large central charge and is predicted to be a strongly chaotic system.\cite{Roberts2015}

\section{Conclusion and Remarks}
\label{sec:con}
In conclusion, we have explored the behavior of
OTO correlators in various models that have a MBL phase. The OTO correlator has recently been used to study black hole dynamics, and has been used to characterize the speed of operator spreading.  In this paper, we show that in the MBL phase the thermally averaged OTO correlators converge to their long time limit by quasiperiodic oscillations within an envelope that decays as a power of time.
We show this analytically within the effective model of  the MBL phase.  The power law decay is due to dephasing caused by the exponentially small long-range effective interactions between the localized integrals of motion. This argument is further verified numerically in three different disordered spin-chain models. The exponent of the power law is linearly proportional to the decay length of the effective interactions, which generally depends on the disorder strength and the temperature.  After a sufficiently long time evolution, the thermally averaged OTO correlator eventually saturates to a constant determined by the overlap between the local operators and the localized integrals of motion.

We further extend our calculation to the thermal phases and find that OTO correlators can apparently approach to zero in different ways. In the Heisenberg model under magnetic field, we observe a power law relaxation to zero even deep inside the thermal phase.  For the Floquet spin model, on the other hand, we find that the OTO correlator approaches zero exponentially in time. We attribute this faster decay to the absence of conserved densities in this Floquet system.

The OTO correlator is sensitive to the weak long-range interactions in MBL phases which distinguish them from noninteracting Anderson localized phases.  Recently, there has been several proposals for measuring the OTO correlator experimentally in systems of cold atoms in optical lattices.\cite{Serbyn2014b, Swingle2016_b, Zhu2016, Yao2016}  A possible advantage of the OTO correlator over other quantities like the entanglement entropy is that it might be more readily studied in the lab.  It would be very interesting to observe these power law decays in real experiments.

\begin{acknowledgements}
We thank Xiongjie Yu for early collaboration on numerical calculation. TZ would like to acknowledge Yingfei Gu for introducing his work\cite{gu_local_2016} and a recent random matrix study\cite{cotler_black_2016} of the thermal phase. XC would like to acknowledge useful discussions with David Luitz and Leon Balents. XC was supported by a postdoctoral fellowship from the Gordon and Betty Moore Foundation, under the EPiQS initiative, Grant GBMF-4304, at the Kavli Institute for Theoretical Physics. 
This work was supported in part by  the National Science Foundation under grant numbers DMR-1306011 (TZ) and  DMR-1408713 (XC,EF), at the University of Illinois at Urbana-Champaign. This work made use of the Illinois Campus Cluster, a computing resource that is operated by the Illinois Campus Cluster Program (ICCP) in conjunction with the National Center for Supercomputing Applications (NCSA) and which is supported by funds from the University of Illinois at Urbana-Champaign. 
\end{acknowledgements}

\bibliography{biblio}
\end{document}